\documentclass[11pt,a4paper]{article}

\usepackage{amsmath,amssymb}
\usepackage{graphicx}
\usepackage{rotating}
\usepackage{cancel}
\usepackage{bm}
\usepackage{xcolor}
\usepackage{comment}
\usepackage{cite}
\usepackage{psfrag}

\usepackage{caption}
\usepackage{subcaption}
\usepackage{enumerate}
\graphicspath{{img/}}

\renewcommand\[{\left[}

\def\bra{\langle}
\def\ket{\rangle}
\newcommand{\amp}[2][0]{\langle #2 \lvert \hat{x} \rvert #1 \rangle}

\begin{document}
\numberwithin{equation}{section}
\title{
\vspace{2.5cm} 
\Large{\textbf{Exploring High Multiplicity Amplitudes:\\ The QM Analogue of the Spontaneously Broken Case
\vspace{0.5cm}}}}

\author{Joerg Jaeckel and Sebastian Schenk\\[2ex]
\small{\em Institut f\"ur Theoretische Physik, Universit\"at Heidelberg,} \\
\small{\em Philosophenweg 16, 69120 Heidelberg, Germany}\\[0.8ex]}

\date{}
\maketitle

\begin{abstract}
\noindent
Calculations of high multiplicity Higgs amplitudes exhibit a rapid growth that may signal an end of perturbative behavior or even the need for new physics phenomena.
As a step towards this problem we consider the quantum mechanical equivalent of $1 \to n$ scattering amplitudes in a spontaneously broken $\phi^4$-theory by extending our previous results on the quartic oscillator with a single minimum \cite{Jaeckel:2018ipq} to transitions $\langle n \lvert \hat{x} \rvert 0 \rangle$ in the symmetric double-well potential with quartic coupling $\lambda$.
Using recursive techniques to high order in perturbation theory, we argue that these transitions are of exponential form $\langle n \lvert \hat{x} \rvert 0 \rangle \sim \exp \left( F (\lambda n) / \lambda \right)$ in the limit of large $n$ and $\lambda n$ fixed.
We apply the methods of ``exact perturbation theory" put forward by Serone \textit{et al.}~in~\cite{Serone:2016qog,Serone:2017nmd} to obtain the exponent $F$ and investigate its structure in the regime where tree-level perturbation theory violates unitarity constraints.
We find that the resummed exponent is in agreement with unitarity and rigorous bounds derived by Bachas~\cite{Bachas:1991fd}.
\end{abstract}

\newpage

\section{Introduction}
\label{sec:introduction}
Perturbative\cite{Cornwall:1990hh,Goldberg:1990qk,Brown:1992ay,Voloshin:1992mz,Argyres:1992np,Smith:1992kz,Smith:1992rq} as well as semiclassical\footnote{For a recent review of semiclassical techniques for multiparticle production see \cite{Khoze:2018mey}.}$^{,}$\footnote{Recently the authors of~\cite{Demidov:2018czx} have used classical simulations to study high multiplicity processes.}~\cite{Son:1995wz,Khoze:2017ifq,Khoze:2018kkz} calculations of high multiplicity $1 \to n$ scattering amplitudes in scalar field theories exhibit an extremely rapid growth with the number of final state particles.
The experimentally observed~\cite{Aad:2012tfa,Chatrchyan:2012xdj} existence of a scalar Higgs boson with a mass of $125$~GeV has taken this problem out of the realm of purely field theoretic interest and has provided us with an explicit upper scale $\lesssim 1600$~TeV \cite{Voloshin:1992rr,Jaeckel:2014lya} (possibly significantly smaller), where we may need better calculational techniques or even new physics.

A particularly interesting form of a new phenomenon could be the ``Higgsplosion" and ``Higgspersion" effect recently proposed in~\cite{Khoze:2017tjt}, that could potentially even address the hierarchy problem and provide for interesting phenomenology \cite{Khoze:2017lft,Khoze:2017uga,Khoze:2018bwa}\footnote{For a discussion on the nature of the underlying quantum field theory, in particular, on aspects of localizability and unitarity in a theory featuring ``Higgsplosion" see~\cite{Belyaev:2018mtd,Monin:2018cbi,Khoze:2018qhz}.}.

For this work our aim is, to some extent, more conservative.
We want to shed light on what are the relevant features that give rise to the growth in high multiplicity amplitudes and whether it can be cured by improved calculational techniques.
To this end we consider a very simplistic toy model: quantum mechanics with a quartic potential.
Here, the vacuum transitions $\amp{n}$ correspond to $1 \to n$ scattering amplitudes.

In studying this toy model we should be keenly aware that, due to its $(3+1)$- instead of $(0+1)$-dimensional nature, quantum field theory is subject to additional features and complications that have to be taken into account.
Important examples are the presence of a non-trivial phase space and the possibility of having weakly coupled, spatially separated final states (see~\cite{Khoze:2018mey} for details).
Nevertheless, we think that it can give important insights into those features, such as the quartic potential, or the existence of a single or multiple minima, that are shared between the two theories.
As we find that advanced calculational techniques probably stop the growth of quantum mechanical amplitudes, it in turn focusses efforts to establish the onset of new phenomena on those aspects of quantum field theory that are different from quantum mechanics.

In a recent paper~\cite{Jaeckel:2018ipq} we have provided significant evidence that vacuum transition amplitudes in the anharmonic oscillator with a single-well potential with quartic coupling $\lambda$ take on the exponential form (conjectured in~\cite{Voloshin:1992nu,Khlebnikov:1992af,Libanov:1994ug,Libanov:1995gh,Bezrukov:1995qh,Libanov:1996vq}),
\begin{equation}
	\amp{n} = \exp \left( \frac{1}{\lambda} F \right) \, .
\end{equation}
At tree-level $F$ turns positive beyond a critical value of $\lambda n$.
This indicates a rapid growth of $\amp{n}$ in the double scaling limit $n\to\infty$ with $\lambda n$ fixed.
However, suitably resummed perturbation theory results in the exponent -- often called \textit{holy grail function} -- being negative, $F < 0$, preventing a rapid growth of the amplitude at high energies~\cite{Jaeckel:2018ipq}.

In this work we extend these results to the symmetric double-well potential,
\begin{equation}
	V(x) = -x^2 + \lambda x^4 \, ,
\end{equation}
which is the quantum mechanical analogue of spontaneously broken $\phi^4$-theory.
We therefore realize an important feature that is essential in the case of the Standard Model Higgs.
In particular, our aim is to establish that also in case of the double-well potential the amplitude takes on an exponential form and to find the sign of $F$.
To do so we use recursive relations to compute $F$ to high orders in a perturbative expansion.
We then apply the method of \textit{exact perturbation theory} (EPT) put forward in \cite{Serone:2016qog,Serone:2017nmd}.
Using this we investigate the behavior of $F$ at values of $\lambda n$ beyond the point where tree-level perturbation theory violates unitarity constraints and find strong indications for a restoration of unitarity.

This work is arranged as follows.
In Section \ref{sec:wavefunctions} we start with a brief review of how to reconstruct wave functions and energy levels of a Schroedinger problem using recursive methods to high order in perturbation theory.
These are the main building blocks for computing vacuum transition amplitudes to highly excited states.
Furthermore, we argue that these amplitudes are of exponential form.
Section \ref{sec:ept} is devoted to introducing the concept of exact perturbation theory and applying it to the holy grail function computed before.
In particular, we present a specific example of EPT to obtain the holy grail function associated to the symmetric double-well potential.
Finally, we conclude in Section \ref{sec:conclusions} by giving a brief summary of our results and future perspectives.

\section{Reconstructing Vacuum Transitions}
\label{sec:wavefunctions}

The main building blocks for computing transition amplitudes from the vacuum to an excited state, $\amp{n}$, are the wave functions and their corresponding energy eigenvalues.
That is, we need to find the eigenfunctions of the Schroedinger operator
\begin{equation}
	\left( - \frac{d}{dx^2} + V(x) - E \right) \psi = 0
\label{eq:schroedingerop}
\end{equation}
in the anharmonic oscillator with a symmetric double-well potential,
\begin{equation}
	V(x) = m^2 x^2 + \lambda x^4 \quad \mathrm{with} \quad m^2 <0 \, , \, \lambda > 0 \, .
\end{equation}
An efficient way to find those solutions is to use recursive relations to high order in perturbation theory, first introduced by Bender and Wu in \cite{Bender:1969si,Bender:1990pd}.
A detailed review and application of this approach to transition amplitudes in the anharmonic oscillator with a single-well potential is given in \cite{Jaeckel:2018ipq}.

Yet, there is an issue in applying perturbative techniques to double-well potentials.
In fact, these methods rely on perturbations around the harmonic oscillator solution.
If we naively tried to use them for the double-well potential, we would have to do perturbation theory in an inverted harmonic oscillator background -- with the obvious problems arising from the instability of the potential.
However, we can choose another reference point of the perturbative expansion which is locally harmonic.
In our example a suitable point is one of the two minima \mbox{$x_\pm = \pm \sqrt{-m^2 / 2\lambda}$} of the double-well potential\footnote{Since the two minima are related by parity, both choices are equivalent.}.
Expanding around $x_+$, shifting the coordinate $\tilde{x} = x - x_+$ and subtracting the zero-point energy yields the asymmetric double-well potential
\begin{equation}
	\tilde{V} (\tilde{x}) = \tilde{m}^2 \tilde{x}^2  + 2 \sqrt{\tilde{m}^2} \sqrt{\lambda} \tilde{x}^3 + \lambda \tilde{x}^4 \, ,
\label{eq:doublewell_asymmetric}
\end{equation}
where we defined $\tilde{m}^2 \equiv -2 m^2$.
Due to its positive mass term the potential $\tilde{V} (x)$ is well suited for the perturbative approach we want to pursue.
Calculationally the cost is the introduction of an additional cubic term $\sqrt{\lambda} \tilde{x}^3$.
For convenience we will later set $m^2=-1$, such that the excitations in the minima are of mass $\tilde{m}^2=2$.

Before we continue let us remark that a constant shift in the ground state energy or in the definition of the position operator does not alter the vacuum transition amplitude $\amp{n}$.
In the case of the ground state energy this is obvious as, even in quantum mechanics, only energy differences are relevant, and an additive constant in the Hamiltonian has no effect on those.
Furthermore, $\amp{n}$ involves different energy eigenstates, which are orthogonal to each other.
Therefore, a constant shift in the position operator does not affect the transition amplitude between those states.

Let us now continue, closely following Bender and Wu's approach \cite{Bender:1969si,Bender:1990pd}.
We can reconstruct the wave functions and energy levels order by order\footnote{Note some technical detail here. Since the additional cubic term $\sqrt{\lambda} x^3$ appears with a fractional power of the coupling, we instead define $\Lambda \equiv \sqrt{\lambda}$ and solve the system by integer orders of $\Lambda$.} in the coupling $\lambda$ of the Schroedinger operator associated to the potential $\tilde{V}(x)$.
This standard perturbative ansatz allows us to compute the normalized vacuum transition amplitudes as described in detail in \cite{Jaeckel:2018ipq},
\begin{equation}
	\mathcal{A}_n \equiv \frac{\amp{n}}{\sqrt{\bra n \vert n \ket \bra 0 \vert 0 \ket}} = \int_{\mathbb{R}} dx \, x \bar{\psi}_n \psi_0 \, .
\end{equation}
Similar to the single-well ($m^2 > 0$) these can be factorized into a tree-level and higher order contributions,
\begin{equation}
	\mathcal{A}_n = \mathcal{A}_n^{\mathrm{tree}} \mathcal{A}_\Sigma \, ,
\end{equation}
where the former is given by (cf.~\cite{Brown:1992ay})
\begin{equation}
	\mathcal{A}_n^{\mathrm{tree}} = \sqrt{\frac{n!}{2 \tilde{m}}} \left( \frac{\lambda}{2 \tilde{m}^3} \right)^{\frac{n-1}{2}} \, .
\label{eq:antree}
\end{equation}
In particular, it turns out that the perturbative series of the amplitude is reproduced \textit{exactly} by
\begin{equation}\label{eq:AnSymbolic}
	\mathcal{A}_n = \mathcal{A}_n^{\mathrm{tree}} \exp \left( \frac{1}{\lambda} F_\Sigma \right) \, ,
\end{equation}
where $F_\Sigma$ can be constructed systematically as a series expansion in $1/n$ (cf.~\cite{Jaeckel:2018ipq}),
\begin{equation}
	F_\Sigma \left( \lambda, n \right) = F_0 (\lambda n) + \frac{1}{n} F_1 (\lambda n) + \frac{1}{n^2} F_2 (\lambda n) + \ldots \, .
\label{eq:FSigmaSymbolic}
\end{equation}
Note that, in the coefficient functions $F_i$, the coupling and energy level only appear as the product $\lambda n$, so for convenience we will define the abbreviation
\begin{equation}
	\epsilon = \lambda n \, .
\end{equation}
As we describe in more detail in~\cite{Jaeckel:2018ipq}, the coefficient functions $F_i$ can be obtained by making a polynomial ansatz in $\epsilon$ for each of the $F_{i}$. We can then expand the exponential in \eqref{eq:AnSymbolic} and compare the corresponding coefficients to the power series $\mathcal{A}_\Sigma$.
Remarkably, a finite number of coefficients of $F_\Sigma$ reproduce infinitely many terms of the perturbative series of $\mathcal{A}_\Sigma$ \emph{exactly}, rendering this resummation very powerful.
We have checked this exact replication to order $\mathcal{O} \left( \lambda^{16} \right)$ of $\mathcal{A}_\Sigma$.
 
In order to study the behavior of the vacuum transition as $n \to \infty$, it is useful to also rewrite the tree-level factor in exponential form,
\begin{equation}
	\mathcal{A}_n = \mathcal{A}_n^{\mathrm{tree}} \mathcal{A}_\Sigma \sim \exp \left\{ \frac{1}{\lambda} \left( F^{\mathrm{tree}} + F_\Sigma \right) \right\} \, .
\end{equation}
That is, the total exponent $F$ of the amplitude consists of two contributions, $F = F^{\mathrm{tree}} + F_\Sigma$.

We now want to consider the double scaling limit $n \to \infty$ and $\lambda \to 0$ with $\epsilon = \lambda n$ fixed.
In this regime the sign of $F$ is crucial.
Any $\epsilon$ where $F$ is positive will ultimately lead to an inevitable growth of the amplitudes in the limit where $n \to \infty$.
We are thus interested in the overall sign of the holy grail function for all $\epsilon$.

\begin{figure}[t]
\centering
	\includegraphics[width=0.5\textwidth]{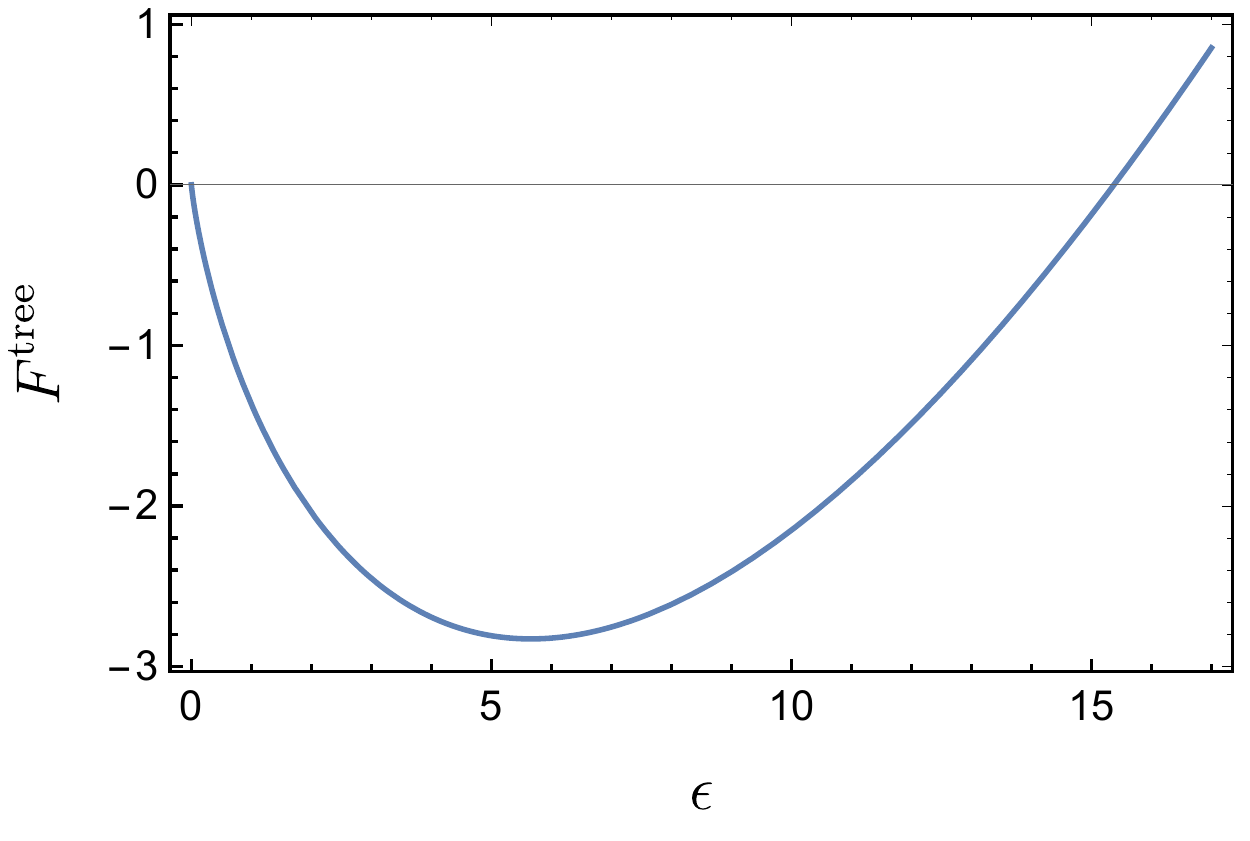}
	\caption{Tree-level contribution to the holy grail function $F$ in the double scaling limit $n \to \infty$, $\epsilon=\lambda n=const$. It exhibits a minimum at $\epsilon=4\sqrt{2}$ and a root at $\epsilon=4\sqrt{2}e$.}
	\label{fig:F_tree}
\end{figure}

In order to establish the sign of $F$, let us discuss its tree-level part first.
Using Stirling's formula for the factorial as $n \to \infty$ in \eqref{eq:antree}, the tree-level contribution can be written as
\begin{equation}
	F^{\mathrm{tree}} \left( \epsilon \right) \sim \frac{\epsilon}{2} \left( \ln \frac{\epsilon}{4 \sqrt{2}} - 1 \right) \, .
\label{eq:FTree}
\end{equation}
This is illustrated in Fig.~\ref{fig:F_tree}.

$F^{\mathrm{tree}}$ indicates a serious issue.
It exhibits a root at $\epsilon = 4 \sqrt{2} e$ where it changes to positive sign, leading to a growth of the amplitude as $n \to \infty$, as pointed out earlier.
The single-well anharmonic oscillator shows a similar behavior of the tree-level contribution, but it turns out that a suitable resummation of $F_\Sigma$ resolves this issue -- $F$ remains negative for any $\epsilon$~\cite{Jaeckel:2018ipq}.
However, such a direct resummation is problematic for the double-well.
In fact, if one computes $F_\Sigma$ in the $1/n$-expansion \eqref{eq:FSigmaSymbolic}, the only contribution that will matter in the double scaling limit is the leading order correction $F_0$, which reads
\begin{equation}
	F_0 \left( \epsilon \right) = \frac{17}{32} \epsilon^2 + \frac{125}{64 \sqrt{2}} \epsilon^3 + \frac{17815}{3072} \epsilon^4 + \frac{87549}{2048 \sqrt{2}} \epsilon^5 + \mathcal{O} \left( \epsilon^6 \right) \, .
\end{equation}

\begin{figure}[t]
\centering
	\includegraphics[width=0.5\textwidth]{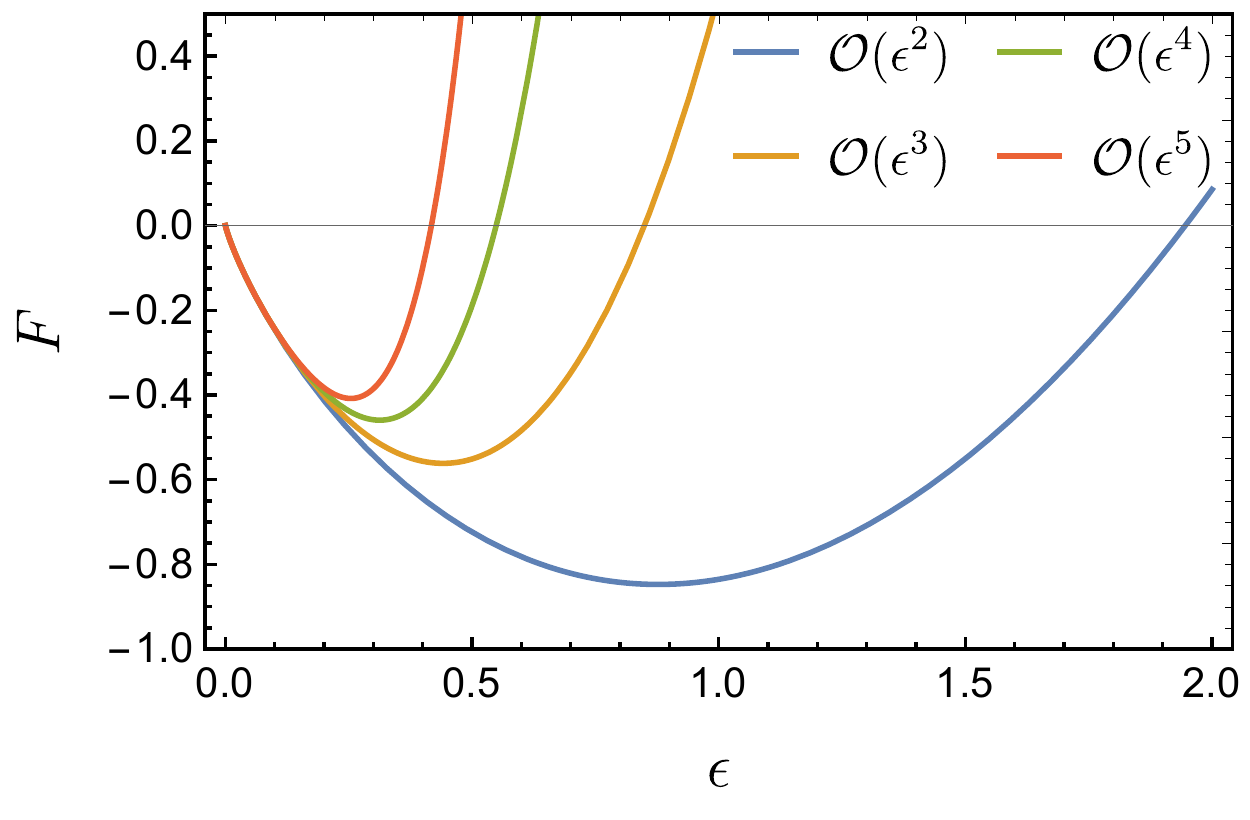}
	\caption{Leading order holy grail function $F = F^{\mathrm{tree}} + F_0$ in the double scaling limit $n \to \infty$ with $\epsilon = \lambda n$ fixed. The label denotes the highest coefficient in $\epsilon$ that is included in the series representation of $F_0$.}
	\label{fig:F_lead}
\end{figure}

The functional form of $F_0$ is strongly governed by the series truncation, such that the root of $F$ is shifted towards smaller values of $\epsilon$ when more terms of the series are included, as can be seen in Fig.~\ref{fig:F_lead}.
In contrast to the single-well the series representation of $F_0$, however, has only positive coefficients instead of alternating ones. This can indicate problems with unambiguous\footnote{These ambiguities can, e.g., be related to poles in the Borel plane, leading to imaginary contributions that are possibly lifted by including non-perturbative effects (see, e.g.~\cite{Jentschura:2004jg,Dunne:2014bca}).} Borel resummation.
Therefore, the sign of $F$ in the case of vacuum transitions in the double-well potential remains unclear.

In summary, applying the same techniques we used for the single-well ($m^2>0$) anharmonic oscillator in \cite{Jaeckel:2018ipq} to a double-well ($m^2<0$) shows that the perturbative expansion of the transition amplitudes is still of exponential form and can in principle be recovered \textit{exactly}.
However, in contrast to the single-well the series expansion of the exponent $F$ for the double-well indicates potential problems with naive Borel resummation, leaving the overall sign unclear.

\section{Exact Perturbation Theory and the Holy Grail Function}
\label{sec:ept}
We have just seen that ordinary perturbation theory provides no clear resolution to the quickly growing high multiplicity amplitudes. Therefore we have to turn to more powerful methods.

The crucial difference to the single-well case is the presence of two degenerate vacua.
This can lead, e.g., to instantons, which cannot be captured by perturbation theory but still have to be included in the quantum mechanical path integral as non-trivial saddles of the action.
It is known that these quantum effects, not respected by our ansatz, can cause perturbative expansions (of e.g.~the vacuum energy) to be non Borel resummable \cite{Brezin:1977gk,Bogomolny:1977ty,Bogomolny:1980ur,Stone:1977au,Achuthan:1988wh,Liang:1995zq}.

Non-perturbative effects (e.g.~instantons) in quantum mechanics and quantum field theory have been extensively studied in the literature, e.g.~\cite{ZinnJustin:1981dx,ZinnJustin:1982td,ZinnJustin:1983nr,Balitsky:1985in,Balitsky:1986qn,Aoyama:1998nt,Aoyama:1997qk,ZinnJustin:2004ib,ZinnJustin:2004cg,Jentschura:2004jg,Marino:2007te,Marino:2008ya,Unsal:2012zj,Aniceto:2013fka}.
However, instead of using instanton calculus we follow a novel approach put forward by Serone \textit{et al.}~in \cite{Serone:2016qog,Serone:2017nmd}.
The principal idea is to recover full non-perturbative results by smart deformations of the perturbative series.
As we will see, this approach is well suited for our consideration, because it makes efficient use of our previous results on the single-well case~\cite{Jaeckel:2018ipq}.
In particular, suitable deformations exploit the non-trivial exponentiation of the amplitude, as they fall in the same class of theories where this powerful resummation is possible\footnote{Nevertheless, it would be very interesting to apply instanton calculus and resurgent trans-series~\cite{Dunne:2014bca} to the example at hand. This could provide a valuable cross-check and potentially also allow us to reach values of $\epsilon$ where the present approach is converging slowly. We leave this for future work.}.

Let us briefly outline the relevant steps of their approach.
In general, we can consider a quantum mechanical potential $V(x;\lambda)$ with coupling $\lambda$ that admits bound states (more precisely $\lim_{\lvert x \rvert \to \infty} V(x;\lambda) = \infty$).
If in addition the potential satisfies $V(x;\lambda) = V(x \sqrt{\lambda}; 1) / \lambda$ it is called \textit{classical}, because the perturbative expansion in $\lambda$ is identical to the expansion in $\hbar$.

Now consider two such classical potentials, $V_0(x;\lambda)$ and $V_1(x;\lambda)$.
The crucial insight is then, that if $V_0(x;\lambda)$ admits a perturbation theory that is Borel resummable to the exact result, the perturbative series of $V(x;\lambda) = V_0(x;\lambda) + \lambda V_1(x;\lambda)$ is also Borel resummable to the exact result, given that $\lim_{\lvert x \rvert \to \infty} V_1(x;1)/V_0(x;1)=0$.
The key is that we treat the part of the potential that causes trouble with Borel summability as a ``small'' perturbation $\sim\lambda$, thereby rearranging the perturbative expansion.
This was coined \textit{exact perturbation theory} in \cite{Serone:2016qog,Serone:2017nmd}.

Each of these two classical potentials can depend on an additional parameter $\lambda_0$.
We can now try to find potentials, depending suitably on $\lambda_{0}$, such that
\begin{equation}
\hat{V}(x;\lambda,\lambda_0) = V_0(x;\lambda,\lambda_0) + \lambda V_1(x;\lambda,\lambda_0)
\end{equation}
and 
\begin{equation}
\hat{V}(x;\lambda,\lambda) = V(x;\lambda),
\end{equation}
i.e.~the original potential is recovered for $\lambda=\lambda_0$.
This will allow us to extract the full information of $V(x;\lambda)$ by a perturbative expansion in $\lambda$  of the \textit{auxiliary} potential $\hat{V}(x;\lambda,\lambda_0)$ and setting $\lambda=\lambda_0$ after performing the Borel resummation.
Serone \textit{et al.}~discuss a variety of quantum mechanical examples in \cite{Serone:2016qog,Serone:2017nmd}.

This method can be useful for potentials with a negative mass term, where standard perturbation theory is not applicable or does not admit an unambiguous Borel resummation.

Let us now apply these ideas to the double-well potential\footnote{The double-well potential is the prime example where instanton solutions play an important role. For instance, they lift the vacuum degeneracy (see, e.g.~\cite{ZinnJustin:2004ib,ZinnJustin:2004cg}).},
\begin{equation}
	V(x;\lambda) = -x^2 + \lambda x^4 \, .
\label{eq:doublewell}
\end{equation}
For this we want to find a potential $\hat{V}(x;\lambda,\lambda_0) = V_0(x;\lambda,\lambda_0) + \lambda V_1(x;\lambda,\lambda_0)$ that reproduces $V(x;\lambda)$ at finite coupling and where $V_0(x;\lambda)$ admits a perturbative expansion that is Borel resummable to the exact result.

Note that the potentials $V_0$ and $V_1$ (and thus $\hat{V}$) are by no means unique.
Even though the final results will be the same after exact resummation, there is a plethora of choices of $V_0$ and $V_1$ which are more or less suited for the approximate computation of certain quantities.
In fact, neglecting constant and linear terms of the potential, the condition on $V_0$ and $V_1$ being \textit{classical} constrains the most general form of $\hat{V}$,
\begin{equation}
	\hat{V}(x;\lambda,\lambda_0) = \left( v_2 + \lambda w_2 \right) x^2 + \left( v_3 + \lambda w_3 \right) \sqrt{\lambda} x^3 + v_4 \lambda x^4 \, ,
\label{eq:vhatgeneral}
\end{equation}
where the coefficients $v_i$ and $w_i$ that belong to $V_0$ and $V_1$, respectively, are functions of $\lambda_0$ only, $v_i = v_i \left( \lambda_0 \right)$ and $w_i = w_i \left( \lambda_0 \right)$.
In order to reproduce the original double-well potential in \eqref{eq:doublewell} at $\lambda_0 = \lambda$ they have to satisfy the conditions
\begin{align}
	 v_2 (\lambda) + \lambda w_2 (\lambda) &= -1 \\
	 v_3 (\lambda) + \lambda w_3 (\lambda) &= 0 \, ,
\end{align}
as well as 
\begin{equation}
v_{4}=1\, .
\end{equation}
This implies that the only free parameters to set up perturbation theory are $v_i ( \lambda_0 )$ (up to additional terms that cancel at $\lambda_0 = \lambda$ in both $v_i$ and $w_i$).
Note, that we have normalized everything to the mass, $m^2=1$.
Furthermore, the $v_i$ have to be chosen such that $V_0$ admits a perturbative expansion that is Borel summable in the end.

While the above conditions yield the most general choice of the potentials $V_0$ and $V_1$, we will focus on a specific example choosing a simple but non-trivial $v_{2}$.

The simplest example presumably constitutes the most intuitive choice of $V_0$ and $V_1$,
\begin{align}
	V_0(x;\lambda,\lambda_0) &= x^2 + \lambda x^4 \\
	V_1(x;\lambda,\lambda_0) &= -\frac{2}{\lambda_{0}} x^2 \, ,
\end{align}
where the single-well anharmonic oscillator potential $V_0(x;\lambda)$ is known to be Borel resummable \cite{Loeffel:1970fe,Graffi:1990pe}.
We then define the potential
\begin{equation}
	\hat{V}(x;\lambda,\lambda_0) = V_0(x;\lambda,\lambda_0) + \lambda V_1(x;\lambda,\lambda_0) = \left(1 - 2\frac{\lambda}{\lambda_0}\right) x^2 + \lambda x^4 \, ,
\label{eq:ex1_hatpotential}
\end{equation}
which reproduces the double-well potential $V(x;\lambda)$ when setting $\lambda_0 = \lambda$,
\begin{equation}
	\hat{V}(x;\lambda,\lambda) = -x^2 + \lambda x^4 \, .
\end{equation}
According to the ideas introduced at the beginning of this section we can now compute any quantity of interest in the double-well $V(x;\lambda)$ by considering the potential $\hat{V}(x;\lambda,\lambda_0)$ instead.
In this potential we can do a perturbative expansion in $\lambda$ (while keeping $\lambda_0$ fixed), perform its Borel resummation and in the end remove the deformation, $\lambda_0 = \lambda$.

\bigskip
In principle we could now plug the new potential into the Schroedinger operator \eqref{eq:schroedingerop} and perform all steps of the original computation of Section~\ref{sec:wavefunctions} in order to obtain $\amp{n}$.
However, in view of the form of $\hat{V}(x; \lambda, \lambda_0)$ in \eqref{eq:ex1_hatpotential} the deformation of the original potential is effectively introducing a mass term that depends on the coupling, $m^2(\lambda) = 1 - 2 \lambda / \lambda_0$.
This allows us to use previous results on $\amp{n}$ obtained in the single-well potential~\cite{Jaeckel:2018ipq}.
In particular, for arbitrary $m^2>0$ we know
\begin{equation}
	\amp{n} = \amp{n}_{\mathrm{tree}} \exp \left( \frac{1}{\lambda} F_\Sigma \right) \, ,
\end{equation}
where the tree-level contribution reads
\begin{equation}
	\amp{n}_{\mathrm{tree}} = \sqrt{\frac{n!}{2m}} \left( \frac{\lambda}{8m^3} \right)^{\frac{n-1}{2}} \, .
\end{equation}
Here $F_\Sigma$ can again be written as a series expansion in $1/n$ (cf.~\eqref{eq:FSigmaSymbolic}).
It is thus dominated\footnote{Note that, in general, a mass term that depends on the coupling of the theory can introduce new factors in the $1/n$-expansion of $F_\Sigma$ which are not subdominant anymore. However, this does not happen in our case. This motivates our choice of deformation.} at large $n$ by $F_0$ which is given by
\begin{equation}
	F_0 (\epsilon) = -\frac{17}{16} \frac{\epsilon^2}{m^3} + \frac{125}{64} \frac{\epsilon^3}{m^6} - \frac{17815}{3072} \frac{\epsilon^4}{m^9} + \frac{87549}{4096} \frac{\epsilon^5}{m^{12}} + \mathcal{O} \left( \epsilon^6 \right) \, .
\end{equation}
We can now plug the mass term $m^2(\lambda) = 1 - 2 \lambda / \lambda_0 = 1 - 2 \epsilon / \epsilon_0$ into $F$ (including tree-level and higher order contributions), do a perturbative expansion in $\lambda$ and rearrange the result in the corresponding $1/n$-expansion.
This yields
\begin{equation}
	\hat{F}_0 (\epsilon, \epsilon_0) = -\frac{17}{16} \epsilon^2 + \frac{125}{64} \epsilon^3 - \frac{17815}{3072} \epsilon^4 + \mathcal{O} \left( \epsilon^5 \right) + \frac{1}{\epsilon_0} \left( \frac{3}{2} \epsilon^2 - \frac{51}{16} \epsilon^3 + \mathcal{O}\left( \epsilon^4 \right) \right) + \mathcal{O} \left( \frac{\epsilon^3}{\epsilon_0^2} \right)\, .
\end{equation}
In principle $\hat{F}_0$ can now be resummed in $\epsilon$ before the deformation of the potential is lifted\footnote{Here one should be careful about the order of the resummation and the lifting of the deformation. If we perform a Borel resummation this requires that we calculate the following Laplace transformation, $F(\epsilon,\epsilon_{0})=\int^{\infty}_{0}dt\, \exp(-t) {\mathcal{B}}F(\epsilon t,\epsilon_{0})$. Here, ${\mathcal{B}}F(\epsilon t,\epsilon_{0})$ is the Borel sum of the power series in $\epsilon t$ while $\epsilon_{0}$ is treated as an external parameter. Note in particular that the argument in the integral is $\epsilon t$ while the external parameter $\epsilon_{0}$ is not multiplied by the integration variable $t$. Only then we can evaluate $F(\epsilon,\epsilon)$.} by inserting $\epsilon_0 = \epsilon$.
This yields the holy grail function $F$ associated to the double-well potential \eqref{eq:doublewell}.
However, by construction we only know a finite number of terms of the series in $\epsilon$.
Thus, we need to make use of an appropriate technique to estimate the behavior from the perturbative coefficients.
Similar to our earlier work on the symmetric case~\cite{Jaeckel:2018ipq}, we first tried to make use of \textit{Pad\'e approximation}.
However, a crucial difference is that here we have to do a separate Pad\'e approximation for every value of $\epsilon_0$ that we want to probe.

While the different Pad\'e approximants appear to converge to negative values for large $\epsilon$, it turns out that the approximation is spoiled by several spurious poles in the small $\epsilon$ region.

Spurious poles in Pad\'e approximants occur also for a number of well-behaved functions and sometimes question the validity of the approximation beyond the pole.
In fact, looking more closely at our series expansion for small $\epsilon_{0}<1$, we find indications for bad behavior. Here, the series appears to forfeit its oscillating sign structure that typically indicates stability in a resummation with a finite number of known terms.
We discuss the technical details of this feature in Appendix~\ref{app:auxparam}. The relevant sign structure can directly be seen from Fig.~\ref{fig:ept_ex1_F0_coeffsign}.

\begin{figure}[t]
\centering
	\includegraphics[width=0.5\textwidth]{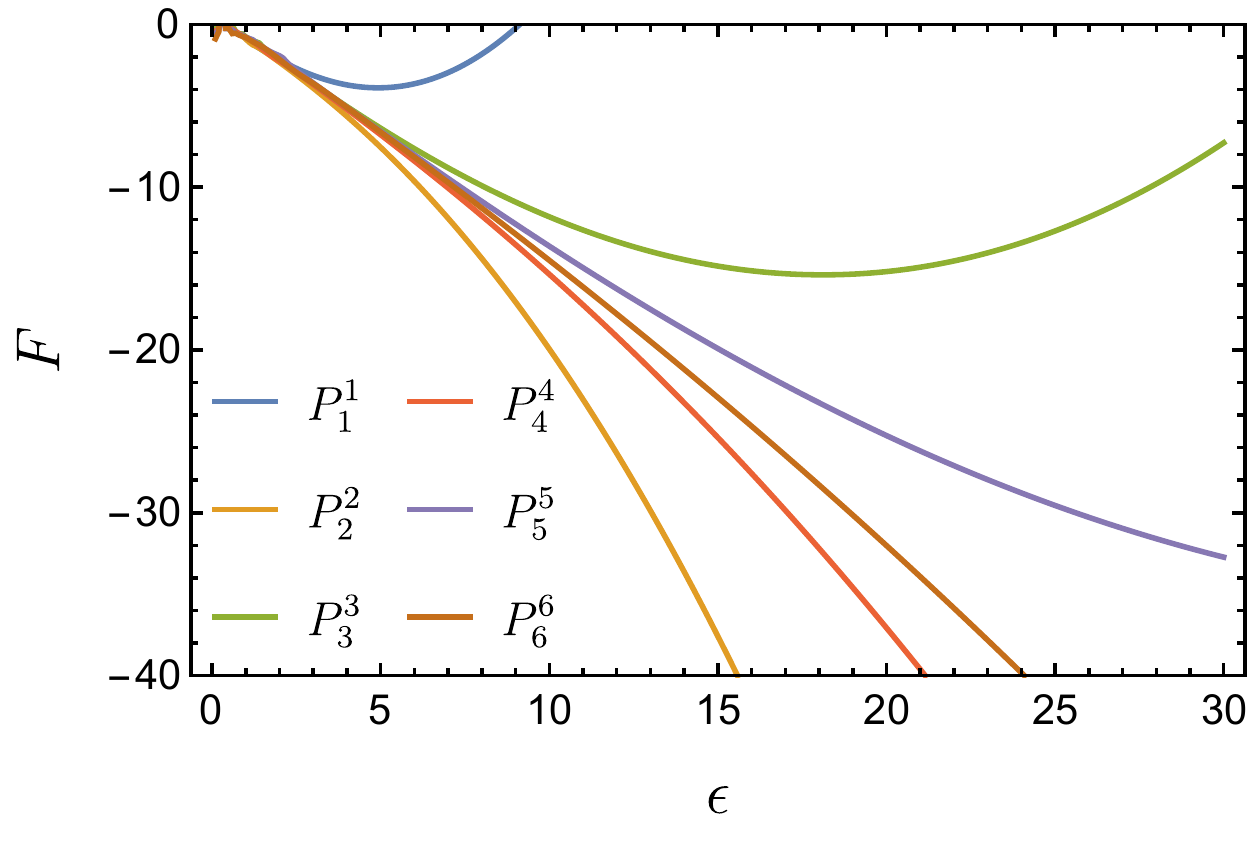}
	\caption{Diagonal Borel-Pad\'e approximants of the holy grail function $F$ in the double scaling limit $n \to \infty, \epsilon = \lambda n=const$. Higher order corrections in the $1/n$-expansion are neglected. $F$ is obtained with the ansatz of exact perturbation theory using the auxiliary potential \eqref{eq:ex1_hatpotential}.}
	\label{fig:ept_ex1_F_borelpade}
\end{figure}

In order to circumvent these problems we have tried different approximation schemes.
Good results are provided by a \textit{Borel-Pad\'e approximation}.
The first few diagonal Borel-Pad\'e approximants of $F$ are illustrated in Fig.~\ref{fig:ept_ex1_F_borelpade}.

We observe that $F$ is indeed negative for a range of $\epsilon$.
In particular, the roots of the resummed $F$ are shifted towards larger $\epsilon$ compared to the naive tree-level contribution \eqref{eq:FTree} when going to higher order in the Borel-Pad\'e approximation.
This gives crucial hints that -- similar to the single-well -- suitably resummed perturbation theory of $F$ resolves the rapid growth $\amp{n}$ for large $n$.

However, we also note that there are still potential problems with the approximation that are more pronounced at small $\epsilon$.
At small values of $\epsilon$ the Pad\'e approximant to the Borel sum inherits the problem of spurious poles due to the all positive signs of the power series expansion.
In the Laplace transformation these poles do not contribute significantly if we take the principle value for the integral. This results in the smooth estimate for the function $F$ shown in Fig.~\ref{fig:ept_ex1_F_borelpade}. The effect of the spurious poles is suppressed at large values of $\epsilon$ since the integrand in the Laplace transform is suppressed exponentially in the region containing the poles. We also remark that such liftable poles are a common feature of Borel-Pad\'e approximations since Pad\'e approximants often feature poles somewhere along the positive real axis.

\begin{figure}[t]
\centering
	\includegraphics[width=0.46\textwidth]{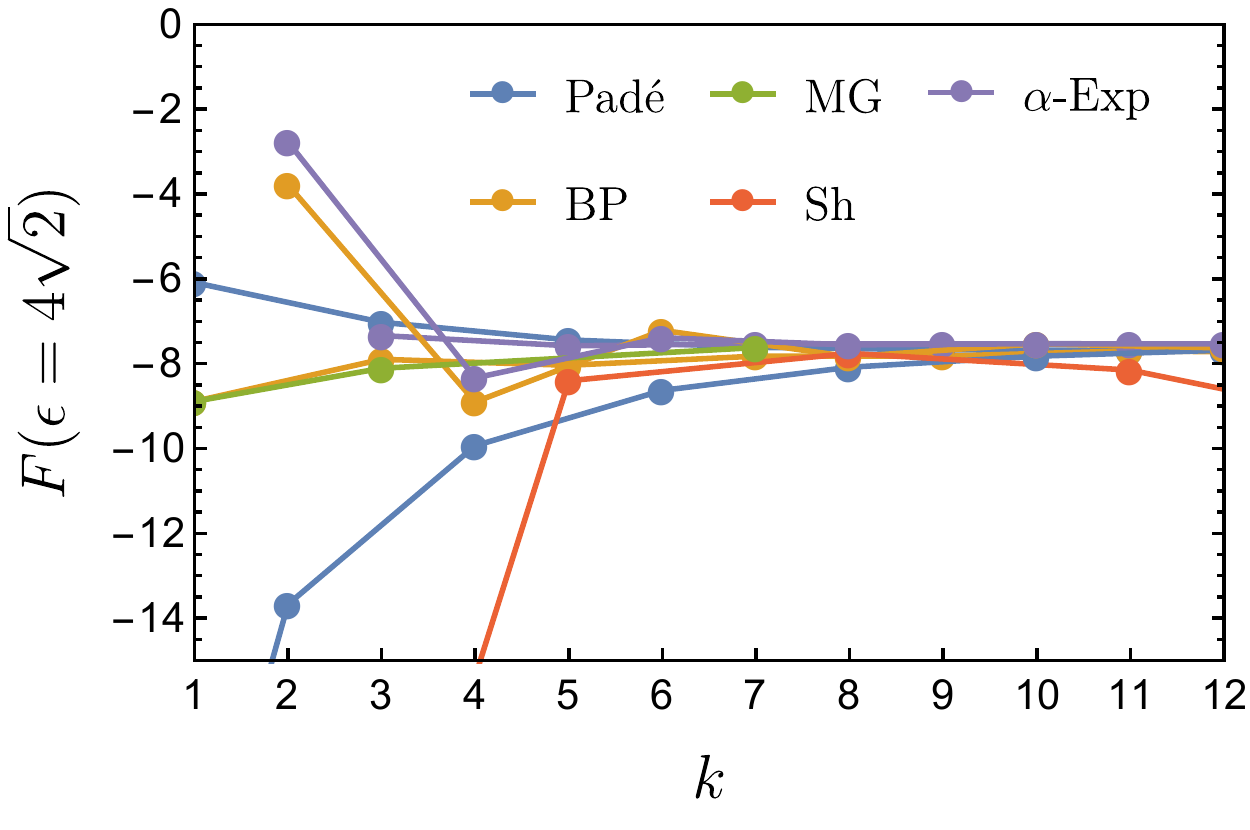}
	\hspace*{0.8cm}
	\includegraphics[width=0.46\textwidth]{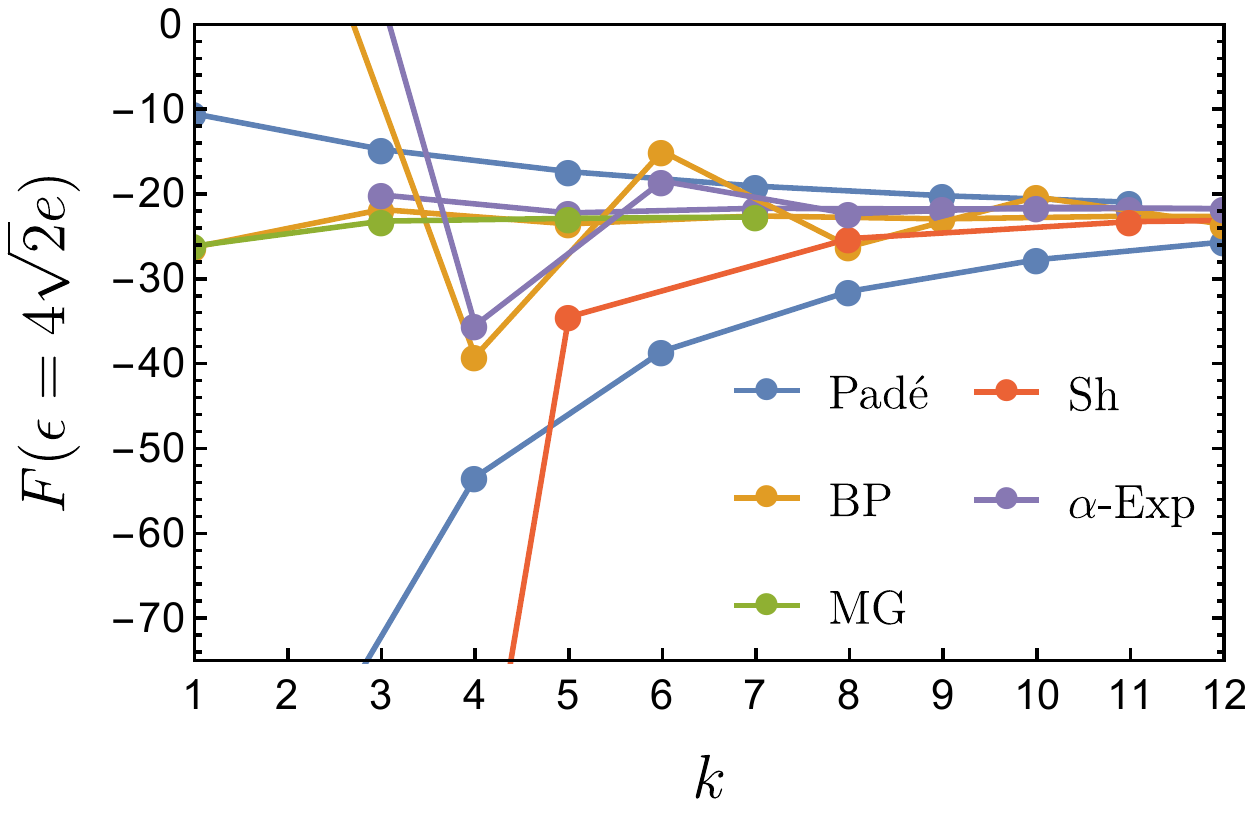}
	\caption{Value of the different approximants of the holy grail function $F$ in the double scaling limit $n \to \infty, \, \epsilon = \lambda n=const$ at the minimum, $\epsilon=4\sqrt{2}$, (left panel) and at the root, $\epsilon=4\sqrt{2} e$, (right panel) of the tree-level holy grail function. The $k$ denotes the number of coefficients of the power series of $F$ taken into account in the corresponding approximation scheme. The different approximation schemes shown are Pad\'e, Borel-Pad\'e (BP), Meijer G (MG) and Shafer (Sh), all of which are discussed in Appendix~\ref{app:approximation}. The label $\alpha$-${\mathrm{Exp}}$ corresponds to the scheme proposed in Appendix~\ref{app:largeorder}.}
	\label{fig:convergence}
\end{figure}

To verify the result of the Borel-Pad\'e approximation, we have tried a number of other resummation schemes.
They are briefly discussed in Appendices~\ref{app:approximation} and \ref{app:largeorder}.
In general, all of them consistently share the same features of negative $F$ at large $\epsilon$ but also some instability.
In Fig.~\ref{fig:convergence} we explicitly illustrate their behavior at the minimum and at the root of the tree-level holy grail function.
The Borel-Pad\'e resummation scheme is shown in yellow.
The other colors correspond to different approximation schemes discussed in Appendices~\ref{app:approximation} and \ref{app:largeorder}. 
While convergence is not completely monotonous for all approximation schemes, they generally agree well with each other.
In particular, at the tree-level zero of the holy grail function the spread between the different results is far smaller than the distance to zero.
This gives a good indication that the sign of the holy grail function is indeed negative at this point.

\begin{figure}[t]
\centering
	\includegraphics[width=0.5\textwidth]{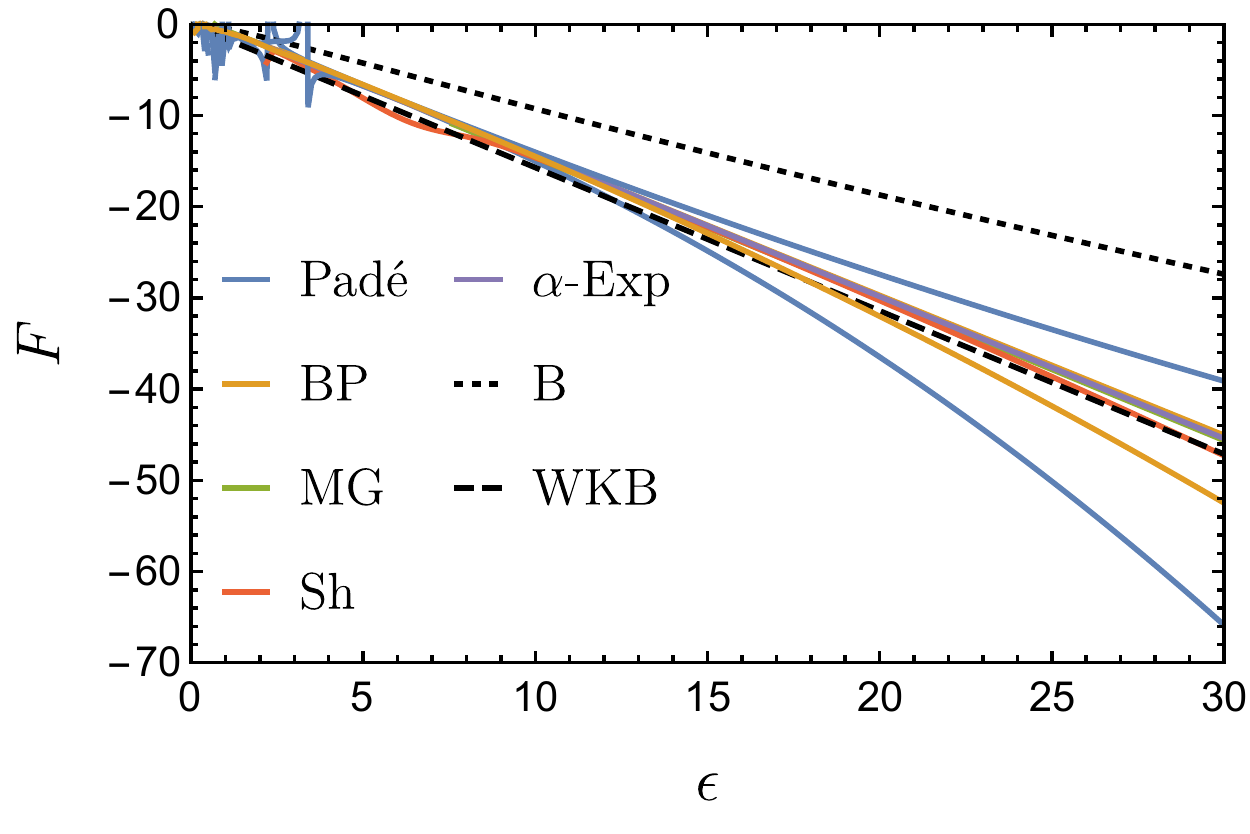}
	\caption{Different approximants (to highest available order) of the holy grail function $F$ in the double scaling limit $n \to \infty, \, \epsilon = \lambda n=const$ (as in Fig.~\ref{fig:ept_ex1_F_borelpade}) compared to WKB estimates \cite{Cornwall:1993rh} and a rigorous bound derived by Bachas \cite{Bachas:1991fd}, labelled WKB and B, respectively. The other labels are as in Fig.~\ref{fig:convergence}.}
	\label{fig:ept_ex1_F_bounds}
\end{figure}

The different approximations of the holy grail function can also be compared to existing results from WKB estimates \cite{Cornwall:1993rh} and a rigorous bound derived by Bachas \cite{Bachas:1991fd}.
They are shown in Fig.~\ref{fig:ept_ex1_F_bounds}.
It turns out that $F$ obtained by EPT is consistent with these results, providing evidence that the ansatz is valid and yields a good approximation to the holy grail function associated to the symmetric double-well potential.

Before concluding let us note that this simple example is just a particular case of a parametrization of $v_2 (\lambda_0) = const$.
In our example we use $v_2=1$.
In Appendix~\ref{app:auxparam} we discuss the convergence as a function of $v_{2}$.

\section{Conclusions}
\label{sec:conclusions}

The behavior of $1 \to n$ scattering amplitudes in $\phi^4$ scalar quantum field theory at high multiplicities remains not well understood.
Calculations of these amplitudes return results that rapidly grow with increasing $n$.
This raises questions about the applicability of the employed calculational techniques, but possibly also about the interpretation of the underlying quantum field theory.
Applied to the Higgs it may even allow for an entirely different phenomenology such as ``Higgsplosion''~\cite{Khoze:2017tjt}.
In order to address these questions we consider the quantum mechanical equivalent, vacuum transition amplitudes to highly excited states, $\amp{n}$, in the anharmonic oscillator with quartic coupling $\lambda$.

We extend our previous work on the single-well potential~\cite{Jaeckel:2018ipq} to the symmetric double-well, that resembles a theory with spontaneous symmetry breaking such as the Higgs sector of the Standard Model.
We find many similarities between both cases.

In particular, using standard perturbation theory to high orders, we find that the amplitude again takes on an exponential form,
\begin{equation}
	\amp{n} = \amp{n}_{\mathrm{tree}} \exp \left( \frac{1}{\lambda} F_\Sigma \right) \, ,
\end{equation}
where the exponent can be constructed in a $1/n$-expansion beyond leading order,
\begin{equation}
	F_\Sigma (\lambda, n) = F_0 (\lambda n) + \frac{F_1(\lambda n)}{n} + \frac{F_2(\lambda n)}{n^2} + \mathcal{O} \left(\frac{1}{n^3}\right) \, .
\end{equation}
As explained in~\cite{Jaeckel:2018ipq} it is non-trivial that the above form reproduces the perturbative series of $\amp{n}$ \textit{exactly}.

However, in the standard perturbative approach we find that $F_\Sigma$ has a series representation with only positive, growing coefficients, such that Borel summability may be problematic.
Consequently, we make use of \textit{exact perturbation theory} \cite{Serone:2016qog,Serone:2017nmd}, a novel approach to (perturbatively) study QM and QFT systems governed by non-perturbative effects.
Considering different resummation schemes, we are able to obtain the holy grail function $F$ associated to vacuum transitions in the double-well.
This suggests that $F$ is indeed negative everywhere in the double scaling limit $n \to \infty$ with $\epsilon = \lambda n = const$,
\begin{equation}
	F(\epsilon) < 0 \quad \forall \epsilon \, .
\end{equation}
That is, the vacuum transitions $\amp{n}$ in the symmetric double-well potential are in line with unitarity bounds for $n \to \infty$.

In summary, our results in this quantum mechanical toy model indicate that -- similar to the single-well potential -- appropriate resummation of the perturbative expansion of vacuum transitions in the symmetric double-well prevents the growth of the amplitude at high multiplicities.
Even though it is just its quantum mechanical analogue, it might still suggest a possible guideline for the resolution of the rapidly growing $1 \to n$ scattering amplitudes in (spontaneously broken) $\phi^4$-theory, including the case of the Standard Model Higgs.

However, we remark that $\phi^4$-theory is a higher dimensional quantum field theory that is subject to additional complications.
Amongst others these include the non-trivial phase space and the existence of weakly coupled, asymptotic states that are not present in quantum mechanics~\cite{Khoze:2018mey}.

It would be very interesting to further investigate if and how our results on quantum mechanics can be extended to quantum field theory, paying special attention to the properties that are unique to higher dimensional theories.

\section*{Acknowledgements}
We would like to thank Valya Khoze and Michael Spannowsky for interesting discussions.
JJ would like to thank the IPPP for hospitality.
SS gratefully acknowledges financial support by the Heidelberg Graduate School of Fundamental Physics.

\appendix
\section{Choice of Auxiliary Potentials}
\label{app:auxparam}

In general the auxiliary potential $\hat{V}(x; \lambda, \lambda_0) = V_0 (x; \lambda, \lambda_0) + \lambda V_1(x; \lambda, \lambda_0)$ deformed by the parameter $\lambda_0$ has to satisfy a few requirements.
For instance, apart from recovering the original potential $V(x; \lambda)$ when the deformation is removed, $\lambda_0 = \lambda$,
\begin{equation}
	\hat{V}(x; \lambda, \lambda) = V (x; \lambda) \, ,
\end{equation}
the potential $V_0$ has to admit bound states and both $V_i$ have to be \textit{classical}, such that their perturbative expansion in $\lambda$ coincides with the expansion in $\hbar$.
Nevertheless, these conditions leave us with a plethora of possibilities to construct $\hat{V}$.

In this section we want to discuss the implications of different choices of the potential $V_0$ and $V_{1}$.
It is intuitive that the physical result after resumming and removing the potential deformation must be independent of the choice of $V_0$ and $V_{1}$.
But, the choice of the deformation can affect the convergence properties of the perturbative expansion and different choices might be useful when calculating different observables.
When deciding which choice is \textit{suitable} one has to deal with several subtleties that we want to discuss in the following.

In order to be in line with the example presented in Section~\ref{sec:ept} we focus on the auxiliary mass term $v_2$ (cf.~\eqref{eq:vhatgeneral}).
In particular, we want to discuss the most simple case $v_2(\lambda_0)=v_2=const$.
Such a parametrization corresponds to the potentials 
\begin{equation}
V_0 = v_2 x^2 + \lambda x^4
\end{equation} 
and 
\begin{equation}
V_1 = -\frac{(v_2+1)}{\lambda_{0}}x^2 \, . 
\end{equation} 
Here we require $v_2 > 0$ such that $V_0$ admits a perturbation theory that is Borel resummable to the exact result.
Combining $V_{0}$ and $V_{1}$ we have the auxiliary potential
\begin{equation}
	\hat{V} \left(x; \lambda, \lambda_0 \right) = \left[ v_2 - \frac{\lambda}{\lambda_0} \left( v_2 + 1 \right) \right] x^2 + \lambda x^4 \, .
\end{equation}
By a suitable rescaling we find that the effective dimensionless coupling of the theory with potential $V_0$ is $\lambda / v_2^{3/2}$.
Thus, different choices of $v_2$ might influence the convergence properties of its associated perturbative expansion.
The case $v_2=1$ corresponds to the simple example considered in \eqref{eq:ex1_hatpotential}.
Let us now consider the behavior as we move away from this point into the two regimes $v_2 \ll 1$ and $v_2 \gg 1$.

In the first case $v_2 \ll 1$ we can see from the effective coupling, $\lambda / v_2^{3/2}$, that already the theory given by $V_0$ is strongly coupled.
Effectively, even at small $\lambda$ we are therefore trying to set up a perturbative expansion that is not well-defined to begin with.

In contrast, perturbation theory for $V_0$ with $v_2 \gg 1$ naively should work well since the effective coupling is very small.
However, we also note that $V_1$ in this case is large, pointing towards potential trouble.
As we will see, this is indeed the case.

In order for EPT to work the perturbative expansion of $F$ has to be Borel resummable at fixed values of both $v_2$ and $\lambda_0$.
In general an alternating sign at high orders (i.e.~up to a finite number of exceptions) indicates well-behaved Borel summability.
A problem with this criterion arises, because by construction we only know a finite number of terms of the perturbative series of $F$.

\begin{figure}[t]
\centering
	\includegraphics[width=0.46\textwidth]{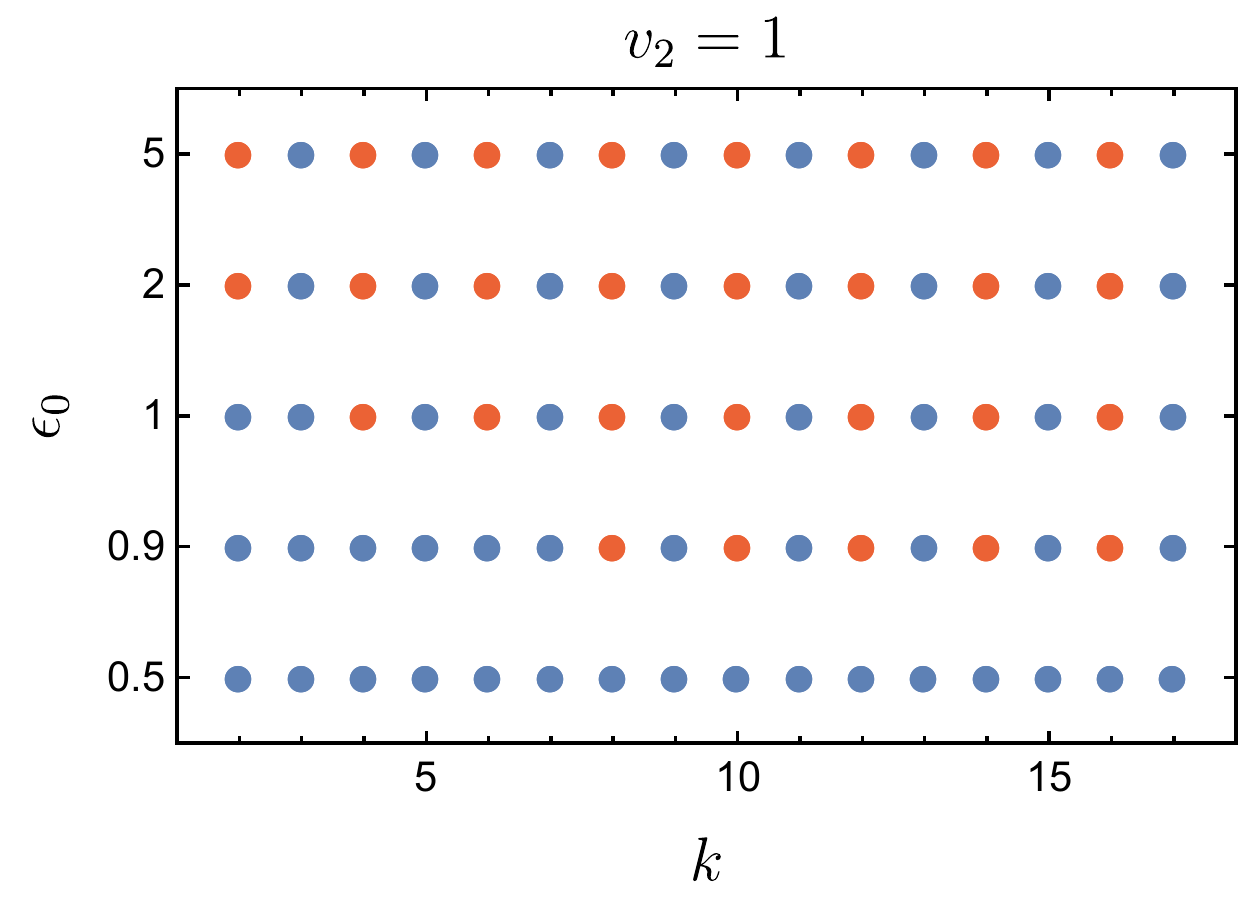}
\hspace*{0.8cm}
	\includegraphics[width=0.46\textwidth]{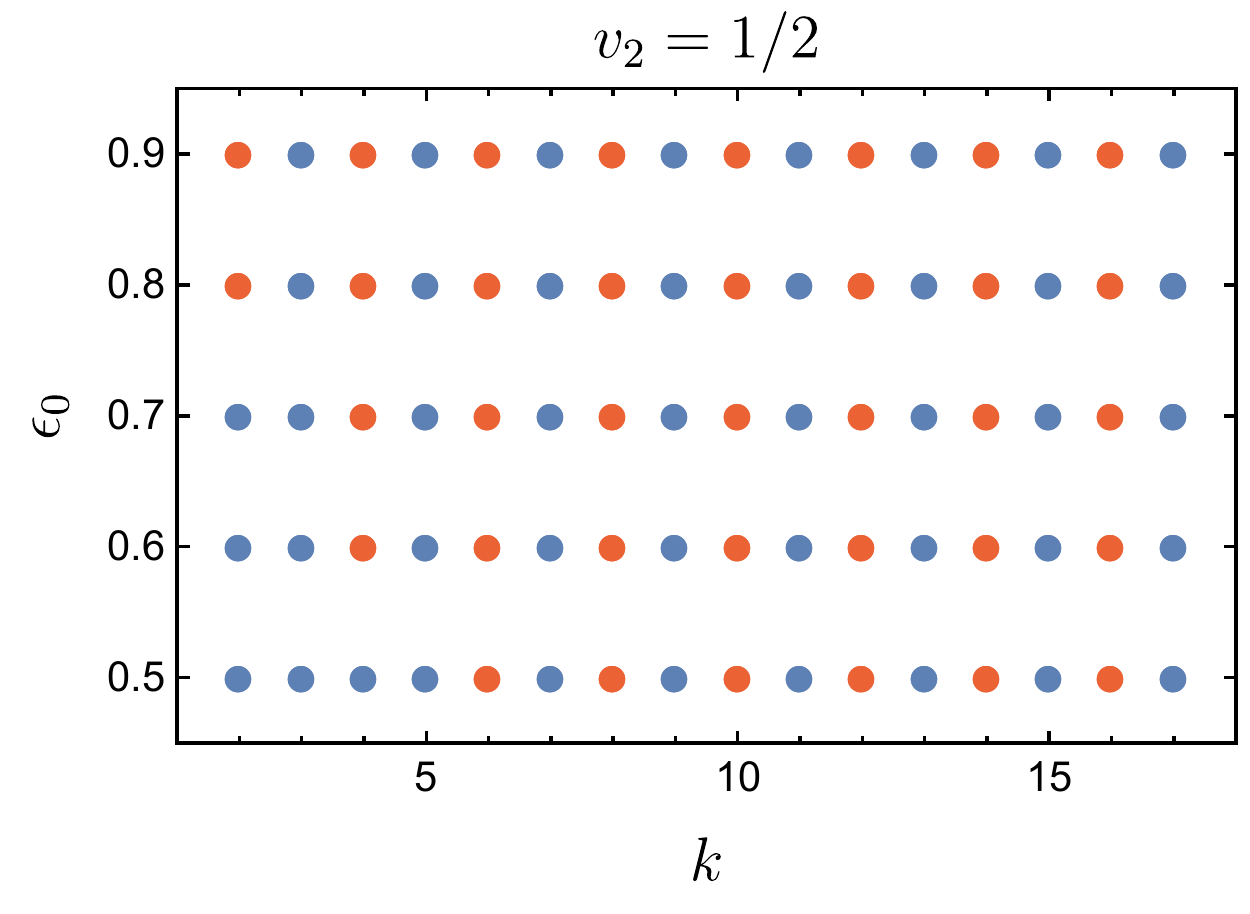}
	\caption{Sign of each coefficient of $\hat{F}_0 (\epsilon,\epsilon_{0}) = \sum_k \hat{F}_{0,k}(\epsilon_{0}) \epsilon^k$ for different values of $\epsilon_0$ shown on the vertical axes. Blue dots denote a positive while red dots denote a negative sign, respectively. The left panel corresponds to $v_{2}=1$ whereas the right panel is for $v_{2}=1/2$. All input parameters are normalized to the mass, $m^2=1$. Note the different scales for $\epsilon_{0}$.}
	\label{fig:ept_ex1_F0_coeffsign}
\end{figure}

Fig.~\ref{fig:ept_ex1_F0_coeffsign} illustrates the signs of the $k$-th coefficient of the series $\hat{F}_0 (\epsilon,\epsilon_{0}) = \sum_k \hat{F}_{0,k}(\epsilon_{0}) \epsilon^k$ for various values of $\epsilon_0$ and two values of $v_2$.
For $v_{2}=1$ (left panel) we can see that a fully alternating series occurs only for values of $\epsilon_{0}>1$.
Thus, after setting $\epsilon_{0}=\epsilon$ we can expect good convergence with the first few approximants only for $\epsilon>1$. 

When lowering $v_{2}$ the alternating sign pattern is preserved for smaller values of $\epsilon_{0}$.
However, as argued above, in this case the effective coupling is larger and despite the alternating sign we need more terms of the series for good convergence.

However, increasing $v_{2}$ too much is problematic, too. In this case the alternating sign pattern only appears for large values of $\epsilon_{0}$.
While the alternating sign will be restored at higher orders, it is clear that approximations based only on the first few terms cannot capture this and behave as if the theory were not Borel summable. 
In this sense such a choice of $v_{2}$ is expected to exhibit worse convergence properties.

In summary, we note that the choice $v_2 \ll 1$ is problematic because we start with a strongly coupled theory, while $v_2 \gg 1$ suffers from an apparent breakdown of Borel summability.
For a reasonable range of $\epsilon$ the choice $v_2 \simeq \mathcal{O}(1)$ seems suitable.
An optimal choice will likely depend on the desired range of $\epsilon$.

\section{Standard Approximation Schemes}
\label{app:approximation}

An essential point of exact perturbation theory is the resummation of a (divergent) power series expansion, before the deformation of the potential is lifted.
In the following we give a brief overview of the different approximation schemes shown in Section~\ref{sec:ept}.

All of them are designed to cope with situations where only a finite number of terms of a power series expansion is known.
Let us consider the formal power series $Z(g) = \sum_{k=0}^\infty z_k g^k$ in the following.

\begin{paragraph}{Pad\'e approximation}
Pad\'e approximation \cite{Pade:1892} is probably the most widely used technique to resum divergent series expansions where only a finite number of terms is available.
Its key idea is to approximate $Z(g)$ by a rational function constructed out of two polynomials $P_M(g)$ and $Q_N(g)$ of degree $M$ and $N$, respectively, such that their ratio coincides with the first few coefficients of $Z$,
\begin{equation}
	P_M + Q_N Z(g) = \mathcal{O} \left(g^{M+N+1} \right) \, .
\end{equation}
The Pad\'e approximant of order $[M,N]$, $Z_{[M,N]}$, is then defined by the condition
\begin{equation}
	P_M + Q_N Z_{[M,N]}(g) = 0 \, .
\end{equation}

It is empirically known that in most examples where Pad\'e approximation is applicable, the diagonal sequence of approximants, $Z_{[N,N]}$, exhibits the best convergence properties to reconstruct $Z$,
\begin{equation}
	Z_{[N,N]}(g) \to Z(g) \quad (N \to \infty) \, .
\end{equation}
In particular, if the coefficients of the power series giving rise to the Pad\'e approximants have an oscillating sign, the true value of $Z$ will typically\footnote{Mathematically a sufficient condition is that the approximated function is a Stieltjes function (see, e.g.~\cite{Bender:1999}). However, in our case this cannot be rigorously deduced from a finite number of coefficients.} lie in between the neighbouring approximants, $Z_{[N,N]}$ and $Z_{[N,N+1]}$.

\end{paragraph}
\begin{paragraph}{Borel-Pad\'e approximation}
Borel resummation \cite{Borel:1899} is typically used, if the large order asymptotics of the coefficients of $Z(g)$ are known.
The method relies on the idea to cancel potential factorial growth of the coefficients by considering the Borel transform of $Z(g)$,
\begin{equation}
	\mathcal{B}Z(g) = \sum_{k=0} \frac{z_k}{k!} g^k \, ,
\end{equation}
hoping for $\mathcal{B}Z$ to converge.
In the end, the factorial factor can be reintroduced by a Laplace transform, such that $Z(g)$ is recovered,
\begin{equation}
	Z(g) = \int_0^\infty dt \, e^{-t} \mathcal{B}Z(gt) \, .
\end{equation}

For all practical purposes, the Borel-Pad\'e approximation now takes into account the fact that the large order asymptotics of $Z(g)$ might not be known.
Instead of computing the Borel transform $\mathcal{B}Z$ exactly, one can try to reconstruct it by a Pad\'e approximant (cf.~previous paragraph), $\mathcal{B}Z_{[M,N]}$.
One can then show that for instance the diagonal Pad\'e sequence will converge to the exact result,
\begin{equation}
	\int_0^\infty dt \, e^{-t} \mathcal{B}Z_{[N,N]}(gt) \to Z(g) \quad (N \to \infty) \, . 
\end{equation}

In this sense Borel-Pad\'e approximation literally combines the Pad\'e approximation with a conventional Borel resummation technique.
\end{paragraph}
\begin{paragraph}{Shafer approximation}
The Shafer approximation~\cite{Shafer:1974} can be understood as the quadratic extension of Pad\'e approximation.
That is, one tries to construct polynomials $P_L(g)$, $Q_M(g)$ and $R_N(g)$ of degree $L, M, N$, respectively, such that
\begin{equation}
	P_L + Q_M Z(g) + R_N Z^2(g) = \mathcal{O} \left( x^{L+M+N+2} \right) \, .
\end{equation}
The Shafer approximant of order $[L,M,N]$, $Z_{[L,M,N]}$, is then defined by the quadratic equation
\begin{equation}
	P_L + Q_M Z_{[L,M,N]}(g) + R_N Z_{[L,M,N]}^2(g) = 0 \, .
\end{equation}
Similar to Pad\'e approximation, the diagonal Shafer approximants $Z_{[N,N,N]}$ will typically have the best convergence properties to $Z$,
\begin{equation}
	Z_{[N,N,N]}(g) \to Z(g) \quad (N \to \infty) \, .
\end{equation}

\end{paragraph}
\begin{paragraph}{Meijer G approximation}
The recently proposed Meijer G approximation scheme~\cite{Mera:2018qte} relies on a similar idea that the Borel-Pad\'e approximation makes use of.
However instead of reconstructing the Borel transform $\mathcal{B}Z(g)$ by means of Pad\'e approximants, it tries to ``guess" its large order asymptotics by representing it by generalized hypergeometric functions,
\begin{equation}
	\mathcal{B}Z (g) \sim {}_{N+1}F_{N} \left( \mathbf{x}, \mathbf{y}; g \right) \, ,
\end{equation}
where the argument vectors $\mathbf{x}$ and $\mathbf{y}$ are defined by singular points of Pad\'e approximants of the successive coefficient ratios of $\mathcal{B}Z(g)$.
The Laplace transform of $\mathcal{B}Z(g)$ can be carried out analytically and yields a Meijer G function.
For details we refer the reader to~\cite{Mera:2018qte}.
\end{paragraph}

\section{Approximation by Guessing the Large Order Behavior of the Borel Sum}
\label{app:largeorder}
Including the large order behavior as done in the Meijer G approximation scheme seems very promising. However, explicitly calculating the ratios between the coefficients in the Borel series for our case we find that they are not well approximated by a constant. Instead they seem to behave approximately as,
\begin{equation}
r_{n}=\frac{a_{n+1}}{a_{n}}\sim f(\epsilon_{0})n^{\alpha},
\end{equation}
with $\alpha$ in the range $1/2-1$ and the function $f$ depending on $\alpha$.

Using this the large order behavior can be accounted for with a so-called $\alpha$-exponential,
\begin{equation}
	 \exp_{\alpha}\left[f(\epsilon_0) \epsilon \right] =  \sum_{k=0}^{\infty} \frac{f \left( \epsilon_0 \right)^k}{k!^{\alpha}} \epsilon^k \quad \mathrm{with} \quad 0 < \alpha \leq 1 \, .
\label{eq:BF_largeorder}
\end{equation}
By fitting to the known coefficients the function $f$ can be accurately expressed in an expansion of $1 / \epsilon_0$.

This function by itself does not very well represent the low order behavior. To include this we can correct the Borel transform by explicitly including the known coefficients up to $n_{\rm max}=14$,
\begin{equation}
	\mathcal{B}F(\epsilon, \epsilon_0) = \sum_{k=0}^{\infty} \frac{f \left( \epsilon_0 \right)^k}{k!^{\alpha}} \epsilon^k + \sum_{k=0}^{n_{\rm max}} \left( \frac{F_{0,k} (\epsilon_0)}{k!} - \frac{f \left( \epsilon_0 \right)^k}{k!^{\alpha}} \right) \epsilon^{k} \, .
\label{eq:BF_fullinformation}
\end{equation}

In principle this could now be directly Laplace transformed, with respect to $\epsilon$ (keeping $\epsilon_{0}$ fixed). However, the behavior can be significantly improved by applying a Pad\'e approximation to the remainder function given by the second part on the right hand side of Eq.~\eqref{eq:BF_fullinformation}. For practical purposes we apply the same order of Pad\'e approximation to the remainder as well as the $\alpha$-exponential.

\begin{figure}[t]
\centering
	\includegraphics[width=0.5\textwidth]{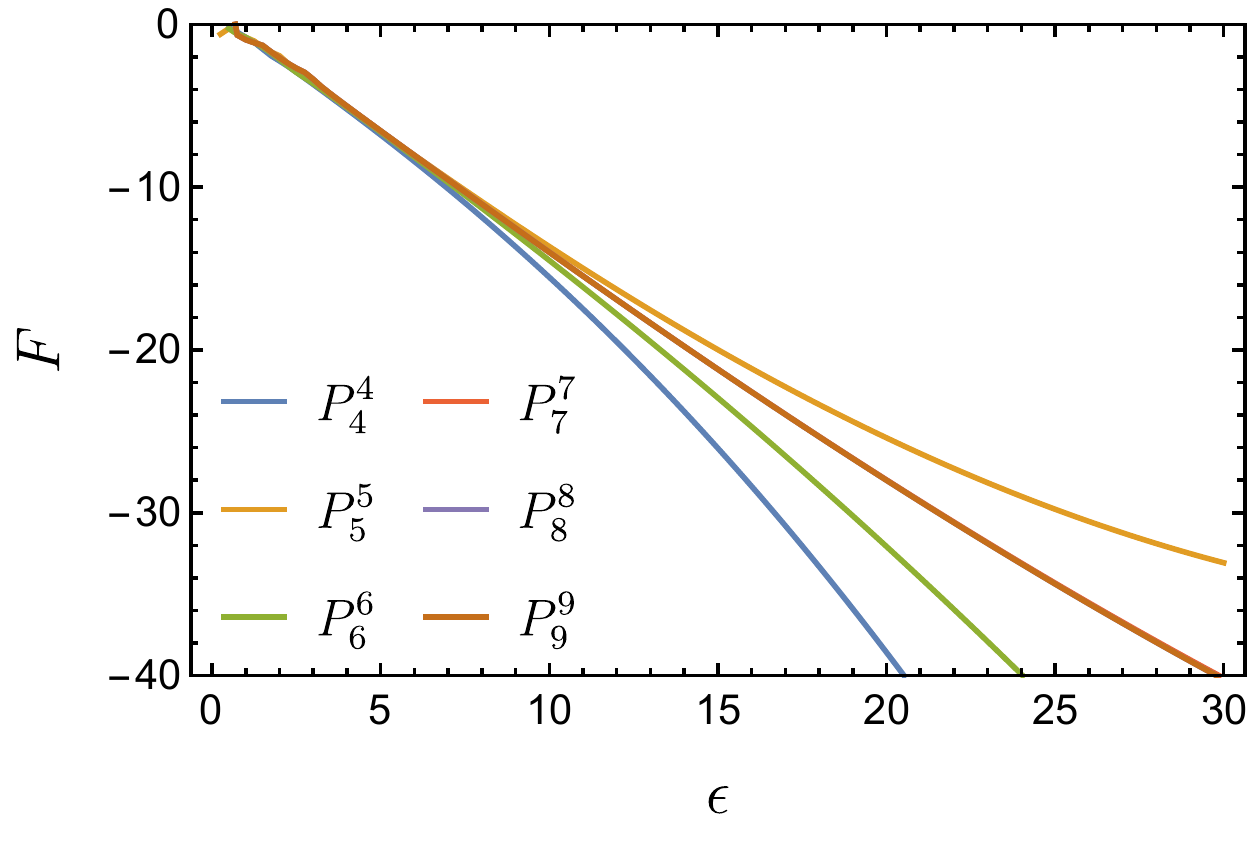}
	\caption{Diagonal Borel-Pad\'e approximants of the holy grail function $F$ with an estimated large order behavior as discussed in this Appendix using Eq.~\eqref{eq:BF_fullinformation}. We can see that the higher approximants are already nicely converged. Note, however, that this does not represent the full error. There is an additional systematic uncertainty at small $\epsilon$ due to the use of the approximate formula~\eqref{largeapp}, as well as a general uncertainty due to our guessing of the large order behavior.}
	\label{fig:ept_ex1_F_borelpade_largeorder}
\end{figure}

We can now apply this to our problem at hand. Fitting the known coefficients for the remainder function we find,
\begin{equation}
\label{largeapp}
f(\epsilon_0) \approx -1.476 + 0.66 / \epsilon_0 + 0.064 / \epsilon_0^2 \quad{\rm for}\quad\alpha=1/2\,.
\end{equation}
The results are shown in Fig.~\ref{fig:ept_ex1_F_borelpade_largeorder}. Note, that this approximation is only good for reasonably large $\epsilon_{0}\gtrsim 2$. Hence, the spread in the different Pad\'e orders possibly underestimates the true uncertainty.

Finally let us finish this discussion with a few words of caution.
By estimating the large order behavior of $\mathcal{B}F$ from the low order coefficients, we gain access to Borel-Pad\'e approximants of higher order.
However, there is a subtle issue when summing $\mathcal{B}F$ that has to be treated with care.
First of all, even though the $\alpha$-exponential has an infinite radius of convergence, $\mathcal{B}F$ cannot be summed naively, i.e.~term by term, as it still contains the low order coefficients $F_{0,k}$.
That is, the remainder function is a finite polynomial. This will dominate over the quickly falling $\alpha$-exponential above a certain critical $\epsilon$, most likely yielding the wrong asymptotics of the Borel-transformed series.
This also means, that the Laplace transform of $\mathcal{B}F$ then does not give a good approximation of the desired function $F$.
This is why we think that the Pad\'e approximation yields a better estimate to the large $\epsilon$ asymptotics of the remainder function.
We have checked that it in fact does not exceed the $\alpha$-exponential part for a wide range of $\epsilon\gtrsim 2$, but still gives a significant contribution to the Laplace transform.

\bibliographystyle{h-physrev}

\end{document}